\documentclass[pra,twocolumn,epsfig,rotate,superscriptaddress,showpacs]{revtex4}
\usepackage{graphicx}
\usepackage{epstopdf}
\usepackage{amsfonts}
\usepackage{amssymb}
\usepackage{amsmath}
\usepackage{subfigure}
\usepackage{prettyref}
\usepackage{float}
\usepackage{amsmath}
\usepackage{amssymb}
\usepackage{esint}

\begin{document}
\title
{Perfect Zitterbewegung oscillations in the Kitaev chain system}

\author{Qi Zhang}
\affiliation{Wilczek Quantum Center and College of Science, Zhejiang University of Technology,
Hangzhou 310014, People's Republic of China}
\author{Jiangbin Gong}
\affiliation{Department of Physics, National University of Singapore, 117542,
Singapore}

\date{\today}
\begin{abstract}
Superconducting systems such as those modeled by the Kitaev Hamiltonian are found to exhibit the Zitterbewegung (ZB) oscillations. Remarkably, the dispersion relation in one-dimensional Kitaev systems allows for wavepackets of arbitrary size undergoing non-decaying ZB without any distortion, with the typical ZB amplitude being one lattice site.  To motivate possible experimental interest in this dynamical aspect of superconducting systems, we further show that certain on-resonance modulation of the Hamiltonian parameter can be exploited to convert ZB oscillations to net drifting of particle's wavepacket and hole's wavepacket along opposite directions, leading to
long-distance particle-hole separation.
\end{abstract}

\maketitle

\section{Introduction}

The Zitterbewegung (ZB) oscillations originally
refers to the jittering motion of free relativistic Dirac particles, as predicted by the Dirac equation \cite{ZB}. However,  a direct experimental detection of ZB is hardly possible due to its extremely high frequency and small amplitude. Furthermore, as far as the ZB dynamics of a quantum wavepacket is concerned, different momentum components of a wavepacket typically involve different ZB frequencies and as such the overall oscillation in the wavepacket position expectation value dephases rapidly.  This makes the observation of coherent ZB oscillations even more challenging. Indeed,  to explore ZB-related quantum dynamics, researchers have actively studied Dirac-like systems with spin-orbit couplings (SOC), both theoretically and experimentally. The studied systems include   single trapped ion \cite{trappedon1,trappedon2}, ultracold atoms \cite{cold1,cold2,cold3,cold4}, band electrons in
graphene \cite{graphene}, as well as cavity electrodynamics \cite{cavitye}.

In this short paper, we propose to explore a rather alternative version of ZB in a type of superconducting systems, where particles and holes are coupled via the Cooper pair mechanism.  For all the above-mentioned ZB studies,
synthesizing the SOC constitutes the starting point; whereas in our consideration below, the ZB is made possible by a pseudo-SOC afforded by the Cooper pair mechanism. As shown below, the pseudo spin degree of freedom is actually the particle state or the hole state, and the coupling is between states of opposite momentum.  Because of this coupling, the obtained ZB can be qualitatively different from previously known cases.

Though our general ideas apply to various superconductor models (so long as momentum-dependent Cooper pairs are present), we choose to use the Kitaev model as a case study. The Kitaev model is the simplest model that realizes Majorana zero-energy state \cite{Kitaev} (Majorana fermion) that has become one important topic in condensed matter physics \cite{Majorana} due to its potential application in fault-tolerant topological quantum computations. Experimentally, the Kitaev system can be realized by a nanowire that has strong SOC (e.g., InSb and InAs nanowire) with s-wave superconductor and a Zeeman field \cite{realization}.  Specifically, by working on one-dimensional Kitaev systems, we show that the dispersion relation of such superconducting systems allows for wavepackets undergoing non-decaying ZB without any distortion. we further show that on-resonance modulation of certain Hamiltonian parameters can be exploited to convert ZB wavepacket oscillations to net drifting of particle's wavepacket and hole's wavepacket along opposite directions, leading to
long-distance particle-hole separation as a coherent quantum control phenomenon.

Before closing this introductory section, we would like to mention that in all previous studies of quantum ZB based on SOC, the ZB amplitude turns out to be much smaller than the width of the wavepacket itself.  A qualitative argument indicates that if this were not the case, then the ZB oscillations will have to dephase and damp very rapidly \cite{zhang,cold3} (the only exception seems to be the ZB of a bound Dirac particle studied in Ref.~\cite{revival}, where the ZB amplitude may have revivals).  As we shall see in the following, the Kitaev system is an unprecedented playground for studying ZB in that (i) there is no decay in the ZB oscillation amplitude (that is, the ZB lifetime can be tuned to infinitely long) and (ii) the ZB oscillation can occur for wavepackets of arbitrary size, i.e. the ratio of the ZB amplitude to the width of the wavepacket undergoing ZB can be arbitrarily large.

\section{ZB in the Kitaev chain}

\subsection{General results}
The mechanism of a low-temperature superconducting system is explained by Cooper pairs, as well captured by the conventional and standard Bardeen-Cooper-Schrieffer (BCS) Hamiltonian expressed in the momentum space.  An alternative superconductor model is the Kitaev model, where terms of electronic tunneling and Cooper pair are expressed on a lattice, i.e., in real space. In particular, the $p$-wave-based superconductor of the Kitaev chain is often described by the following Hamiltonian,
\begin{equation} \label{Hamiltonian}
H=-\mu\sum_{j}c_{j}^\dag c_{j}-\sum_{j}(t_p\ c_{j}^\dag c_{j+1}+H.c.)-\sum_{j}(d\ c_{j}^\dag c_{j+1}^\dag+H.c.)
\end{equation}
where $j$ is the lattice coordinate; $t_p$ and $d$ are the tunneling integral and superconducting pairing amplitude for electrons between the nearest neighboring sites. The real $\mu$ parameter is the chemical potential. Though our discussions below apply to two-dimensional Kitaev models as well, we restrict ourselves to the above one-dimensional model.  

When the chain is long enough and our main concern is not the edge states (like Majorana fermion) but the bulk properties, it is convenient to carry out a Fourier transformation to the $k$-space (momentum space), i.e., $c_k^\dag=\frac{1}{\sqrt{N}}\sum_jc_j^\dag \exp{({\rm i}jka_l)}$ to write the Hamiltonian as ($j$ is the lattice coordinate.  the lattice constant $a_l$ is taken to be unity throughout),
\begin{eqnarray} \label{HamiltonianK} \nonumber
H_k&=&\sum_k(\xi(k) c_k^\dag c_k+\Delta(k) c_k^\dag c_{-k}^\dag+\Delta(k)^* c_k c_{-k}) \\
&=&\sum_k\left(\begin{array}{cc}c_k^\dag&c_{-k}\end{array} \right) \left(\begin{array}{cc}\xi(k)&\Delta(k)\\ \Delta(k)^*&-\xi(k)\end{array} \right) \left(\begin{array}{c}c_k\\c_{-k}^\dag\end{array} \right),
\end{eqnarray}
where $\xi(k)=-\mu-2t_p\cos(k)$ and $\Delta(k)={\rm i}2d\sin(k)$.  As seen above, the Kitaev model defined in real space can be equivalently expressed in the momentum space, thereby assuming a form identical with the BCS Hamiltonian.

As in the case of the BCS model, the Kitaev chain defined in Eq.~(\ref{HamiltonianK}) has both ground and excited states at a given fermion energy, which are determined by the number of excited Bogoliubov quasi-particles. Suppose $|E\rangle$ is an eigenstate of $H_k$, then a rather arbitrary spinor-like state,
\begin{equation}
|u,v\rangle=(uc_k^\dag+vc_{-k})|E\rangle
\end{equation}
will not be stationary. Instead, it will evolve under the superconductor Hamiltonian $H_k$. The only exception arises if $u,v$ happen to satisfy the Bogoliubov condition and $(uc_k^\dag+vc_{-k})|E\rangle$ then represents the creation or annihilation of a quasi-particle. Certainly,  the dynamical evolution of $u,v$ must satisfy a two-mode Schr\"odinger equation governed by the effective Hamiltonian,
\begin{equation} \label{effective}
H_{\text{eff}}=\left(\begin{array}{cc}\xi(k)&\Delta(k)\\ \Delta(k)^*&-\xi(k)\end{array} \right).
\end{equation}
Thus a particle state on top of an overall eigenstate with amplitude $|u|^2$ and a hole state of inverse momentum with amplitude $|v|^2$ are coupled together. If the particle-hole representation is understood as the two component of a pseudo spin degree of freem,
then the above effective Hamiltonian $H_{\text{eff}}$ describes a pseudo SOC. This pseudo SOC is different from early SOC model Hamiltonians in that the internal state dynamics is always accompanied by two opposite momentum values rather than the same momentum values. The implication of this coupling will be discussed much later. Here it is worth noting that the periodic dependence of $H_{\text{eff}}$ on $k$ reflects the periodic quasi-momentum nature for discrete lattice model, and the one period interval is just the one-dimensional Brillouin zone.

Without loss of generality, the tunneling parameter $t_p$ and the pair amplitude parameter $d$ are assumed to be real positive numbers.
We now examine the motion of an initial state ($t=0$) as a product
state of a one-dimensional wavepacket on the lattice and an internal
two-component pseudo- ``spinor" state,
\begin{equation} \label{ip}
\langle j|\psi(0)\rangle=G(j)\left(\begin{array}{c}a\\b\end{array}\right),
\end{equation}
where $G(x)$ is a broad Gaussian in real space centered at $j=0$;
the central momentum of the initial wavepacket along the chain is assumed to be zero.
In connection with $H_{\text{eff}}$ defined above, this initial state stands for a superposition of particle and hole states, coherently delocalizing over some lattice sites with a Gaussian profile.

Since ZB can be more clearly investigated in the momentum space, we carry out the Fourier
transformation of the above initial state to arrive at a narrow wavepacket in the
momentum representation, i.e.,
\begin{equation} \label{ipk}
|\psi_k(0)\rangle=\langle
k|\psi(0)\rangle=\left(\begin{array}{c}g(k)a\\g(-k)b\end{array}\right),
\end{equation}
where $g(k)$ or $g(-k)$ is also a Gaussian as the Fourier transformation of $G(j)$.  Here $g(k)=g(-k)$ is an even function of $k$ because the initial Gaussian state is symmetric in both the lattice space and in the momentum space.

Interpreting the momentum-space effective Hamiltonian $H_{\text{eff}}$ as a magnetic Hamiltonian for a (pseudo) spin, one can see that the internal state specified in Eq.~(\ref{ipk}) evolves in the presence of two components of a ``magnetic
fields": one along ``$z$" of strength $\xi(k)$ and the other along ``$y$"
with strength $-2d\sin(k)$. The
total ``magnetic field" strength is $\sqrt{\xi^2+4d^2\sin(k)^2}$
and the direction of the total magnetic field is characterized by an
angle $\arctan[2d\sin(k)/\xi(k)]$.

To gain more insights let us first make an expansion to the first order of $k$, by considering a narrow wavepacket in the momentum space (we stress that later we will drop this kind of approximation) \cite{cold2}.  Keeping the effects up to the first order of $k$, we have
\begin{eqnarray} \label{approximation} \nonumber
\arctan\left(\frac{2d\sin(k)}{-\xi(k)}\right)&\approx& \frac{2d}{\mu+2t_p}k; \\
\sqrt{\xi^2+4d^2\sin(k)^2}&\approx& \mu+2t_p,
\end{eqnarray}
where we have assumed that $\mu+2t_p>0$ and $\mu+2t_p\gg kd$. For the case $\mu+2t_p<0$, we can perform an analogous approximation, which is not repeated here.
Physically, this approximation is
to assume that, for different $k$ components, their effective
Zeeman splitting is almost the same, but with the
internal state precessing around slightly different directions linearly
dependent on $k$.  As shown in the following calculations, this linear $k$-dependence on the magnetic field orientation angle will carry over to the wavepacket dynamics, resulting in a wavepacket phase linearly proportional to $k$. Such a phase linear in $k$ indicates a shift of the wavepacket center in the position space.  Note that this is obtained under the assumption that the total magnetic field strength is approximately independent of $k$, so all the different $k$ components oscillate at the same angular frequency.  Because this common angular frequency is
given by $\omega=2(\mu+2t_p)/\hbar$ (in dimensionless units), the shift in the wavepacket centre should be periodic with a period $T=\frac{2\pi}{\omega}=\frac{\pi}{\mu+2t_p}$.

More specifically,  according to this approximation, the evolution of the initial state specified in (\ref{ipk}) can be explicitly written down. For clarity and concreteness, we write down the time-evolving state at $t=T/2=\frac{\pi}{2(\mu+2t_p)}$ for a specific case $a=-b=1/\sqrt{2}$ (neglecting an overall phase),
\begin{equation} \label{ipkf}
|\psi_k\left(\frac{\pi}{2(\mu+2t_p)}\right)\rangle=\frac{1}{\sqrt{2}}\left(\begin{array}{c}g(k)\\g(-k)
\end{array}\right)e^{-{\rm i}\frac{2kd}{\mu+2t_p}}.
\end{equation}
In above, the first component of the state is for the particle component, depicting a wavepacket on the lattice whose center is located at $j=\frac{2d}{\mu+2t_p}$.  This becomes more obvious if we perform an inverse Fourier transformation of the first component to real space,
arriving at
\begin{equation}
\langle j|\psi_{\text{electron}}\rangle= \frac{1}{\sqrt{2}}G\left(j-\frac{2d}{\mu+2t_p}\right).
  \end{equation}
  The second component in the above expression is for the hole component, depicting a ``hole" wavepacket centered at $j=-\frac{2d}{\mu+2t_p}$. Indeed, the corresponding inverse Fourier transformation yields
\begin{equation}
\langle j |\psi_{\text{hole}}\rangle=\frac{1}{\sqrt{2}}G\left(j+\frac{2d}{\mu+2t_p}\right).
\end{equation}
 Clearly then,  despite the fact that initially the particle and hole wavepackets are spatially on top of each other,  they start to separate from each other due to the ZB oscillations.  The net result at $t=T/2$ is a net separation of particle from hole on the Kitaev chain.

When $t=T=\frac{\pi}{\mu+2t_p}$, the ``spinors" for all momentum components rotate back to the initial state, and as a result the particle's wavepacket once again exactly overlaps with the hole's wavepacket, both centered at the $j=0$ lattice site. This also indicates that both particle and hole undergo opposite ZB oscillations with period $T=\frac{\pi}{\mu+2t_p}$, amplitude $A\sim\frac{2d}{\mu+2t_p}$. Such kind of ZB phenomenon is qualitatively different from the conventional ZB studied so far (that is, in the conventional ZB, all the components of the ``spinor" always undergo shift in the same manner).

For completeness, the time-evolving state in the above case in coordinate space is given as (neglecting an overall phase),
\begin{eqnarray}  \nonumber
&&\langle j|\psi(t)\rangle=\cos[(\mu+2t_p)t]\frac{1}{\sqrt{2}}\left(\begin{array}{c}G(j)\\-G(j)\end{array}\right) \\
&&+\sin[(\mu+2t_p)t]\frac{1}{\sqrt{2}}\left(\begin{array}{c}G\left(j-\frac{2d}{\mu+2t_p}\right) \\ G\left(j+\frac{2d}{\mu+2t_p}\right)\end{array}\right).
\end{eqnarray}

\subsection{perfect ZB oscillations without damping}
The central ZB physics discussed above is still based on the approximation depicted by Eq.~(\ref{approximation}), where we essentially require $\mu+2t_p\gg (\delta k)d$, where $\delta k (\sim\frac{1}{\delta j})$ and $\delta j$ are the characteristic width of the wavepacket in the momentum space and in the position space, respectively.  For this assumption to be valid, the wavepacket should be sufficiently narrow in the momentum space or sufficiently wide in the position space. For this reason,  the ZB amplitude $A$ turns out to be much smaller than the wavepacket's width in the position space, i.e., $A\ll\delta j$ (please refer to Ref.~\cite{zhang} for a detailed analysis and refer to Ref.~\cite{trappedon2} for an experimental study). More importantly, after a few ZB periods, the ZB oscillations start to damp and the wavepacket shape starts to distort once the effects beyond the approximation to the first order of $k$ kick in \cite{cold3}.  For example, soon enough, different $k$-components of the wavepacket start to oscillate at different phases because the effective field strength they experience is after all different to the second order of $k$.

Remarkably, for the Kitaev chain considered here, we find that the above-mentioned reasons to give rise to non-perfect ZB oscillations and wavepacket distortion can become irrelevant altogether. This finding is also the main result of this work. In particular, coming back to the effective Hamiltonian $H_{\text{eff}}$ in Eq.~(\ref{effective}) and noting the trigonometric dependence of $\xi_k$ and $\Delta(k)$, we observe that the following condition
\begin{eqnarray} \label{magiccondition} \nonumber
\mu&=&0;
\\t_p&=&d
\end{eqnarray}
constitutes a magic situation.  Under this parameter choice,  we exactly have
\begin{eqnarray} \label{exact} \nonumber
\arctan\left(\frac{2d\sin(k)}{-\xi(k)}\right)&=& k; \\
\sqrt{\xi^2+4d^2\sin(k)^2}&=& 2d.
\end{eqnarray}
That is, regardless of the value of $k$, the total ``magnetic" field strength is always independent of $k$ and the angular dependence of the field on $k$ is strictly linear. That is,  the above-mentioned first-order approximation (\ref{approximation}) is no longer needed as the very same relations hold {\it precisely} for all $k$ values.  As a matter of fact, the effective Hamiltonian in the momentum space now becomes
\begin{equation} \label{effective2}
H_{\text{eff}}=2d \left(\begin{array}{cc}\cos(k)& {\rm i}\sin(k) \\ - {\rm i} \sin(k)&-\cos(k)\end{array} \right).
\end{equation}
There it can be seen more evidently that for the $k$-component, the pseudo-spin describing particle and hole states experiences a ``magnetic field" pointing at, in the standard notation for a spherical coordinate system, the direction of $(\theta, \phi)$, with $\theta=\mod(k,2\pi)$ up to a $2\pi$ shift, and $\phi=-\pi/2$. The associated field strength is independent of $k$. The $k$-independence of the effective field strength guarantees that the oscillations of different $k$ components are always in phase. According to Eq.~(\ref{ipkf}), in the parameter condition specified in (\ref{magiccondition}), the ZB amplitude is strictly to be (in units of lattice constant $a_l$),
\begin{equation}
A=\frac{2d}{\mu+2t_p}\equiv 1,
\end{equation}
i.e., the ideal ZB amplitude is exactly one lattice site.

With the relation in Eq.~(\ref{exact}) being exact and following the same derivation as before,
one immediately arrives at an ideal ZB oscillations without any damping and deformation of the initial Gaussian wavepacket, where the width of the initial wavepacket can be prepared in arbitrary size. Specifically, in this magic condition an initial state of spatial $\delta$-function profile, i.e., particle and hole states localizing on one lattice site that correspond to plane waves in momentum space, can equally implement the ideal ZB dynamics with the $\delta$-profile kept unchanged. Following exactly the same procedure as before but now with an initial state as a delta-function-profile $\delta(j)$, which behaves as $\delta(j)=1$ for $j=0$ and $\delta(j)=0$ otherwise, we have
\begin{eqnarray} \nonumber
G(j)&=&\delta(j), \\
g(k)&=&\frac{1}{\sqrt{2}}\exp(-k\cdot j_0), \quad \text{with}\quad j_0=0,
\end{eqnarray}
i.e., with the initial state taken as,
\begin{equation}
\langle j|\psi(0)\rangle=\frac{1}{\sqrt{2}}\delta(j)\left(\begin{array}{c}1\\-1\end{array}\right),
\end{equation}
we derive the wavefunction evolving with time as (neglecting an overall phase),
\begin{equation}
\langle j|\psi(t)\rangle=\cos(2td)\left(\begin{array}{c}\frac{\delta(j)}{\sqrt{2}}\\-\frac{\delta(j)}{\sqrt{2}}\end{array}\right)+\sin(2td)\left(\begin{array}{c}\frac{\delta(j-1)}{\sqrt{2}}\\ \frac{\delta(j+1)}{\sqrt{2}}\end{array}\right).
\end{equation}
It can be seen evidently that the occupation probabilities of the two sub-$\delta$-profiles over one lattice site for particle state (first component of spinor) are $\cos^2(2td)/2$ and $\sin^2(2td)/2$, respectively, and those for hole state (second component of spinor) are the same except that the spatial direction is opposite, as illustrated in Fig.~1. As time evolves, these two occupation probabilities oscillate, thus giving rise to a time dependence of the average position for both particle state and hole state as,
\begin{eqnarray} \nonumber
\langle j_{\text{electron}}\rangle&=&0\cdot\frac{\cos^2(2td)}{2}+1\cdot\frac{\sin^2(2td)}{2}=\frac{\sin^2(2td)}{2}\\ \nonumber
\langle j_{\text{hole}}\rangle&=&0\cdot\frac{\cos^2(2td)}{2}-1\cdot\frac{\sin^2(2td)}{2}=-\frac{\sin^2(2td)}{2},
\end{eqnarray}
which is nothing but the ZB phenomenon with ZB amplitude being one lattice constant.

This theoretical prediction is fully verified by Fourier transformation and dynamics simulations. In Fig.~2, the time dependence of the mean positions of particle and hole, as well as the spatial profile of their respective wavepackets are shown.  The ZB oscillations are shown to be perfect. Though not shown on the same figure, further numerical results confirm that this type of ZB oscillations will not degrade at all times, including the case of ZB amplitude comparable to the wavepacket width and the case of wavepacket coherently delocalizing over a great number of lattice sites. However, in any case, the ZB amplitude is exactly one lattice constant.
\begin{figure}[t]
\begin{center}
\vspace*{-0.8cm}
\par
\resizebox *{8cm}{6cm}{\includegraphics*{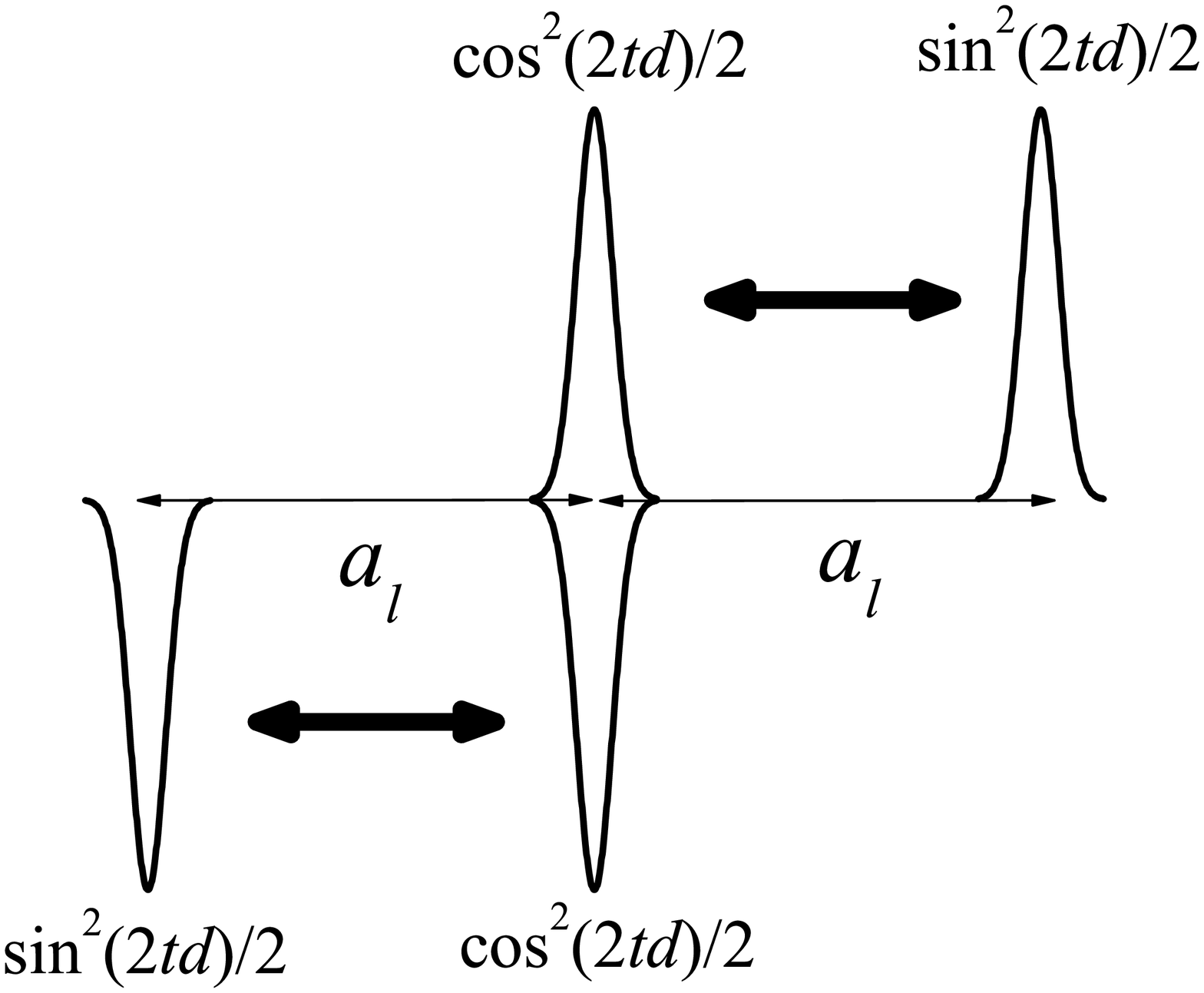}}
\end{center}
\par
\vspace*{-0.5cm}  \caption{ Schematic plot of on-site electron-hole wavepacket motion satisfying the Kitaev
equation. Cyclic population transfers associated with electron and hole components between neighbouring sites
give rise to ZB oscillation.}
\end{figure}

\begin{figure}[t]
\begin{center}
\vspace*{-0.8cm}
\par
\resizebox *{8cm}{6cm}{\includegraphics*{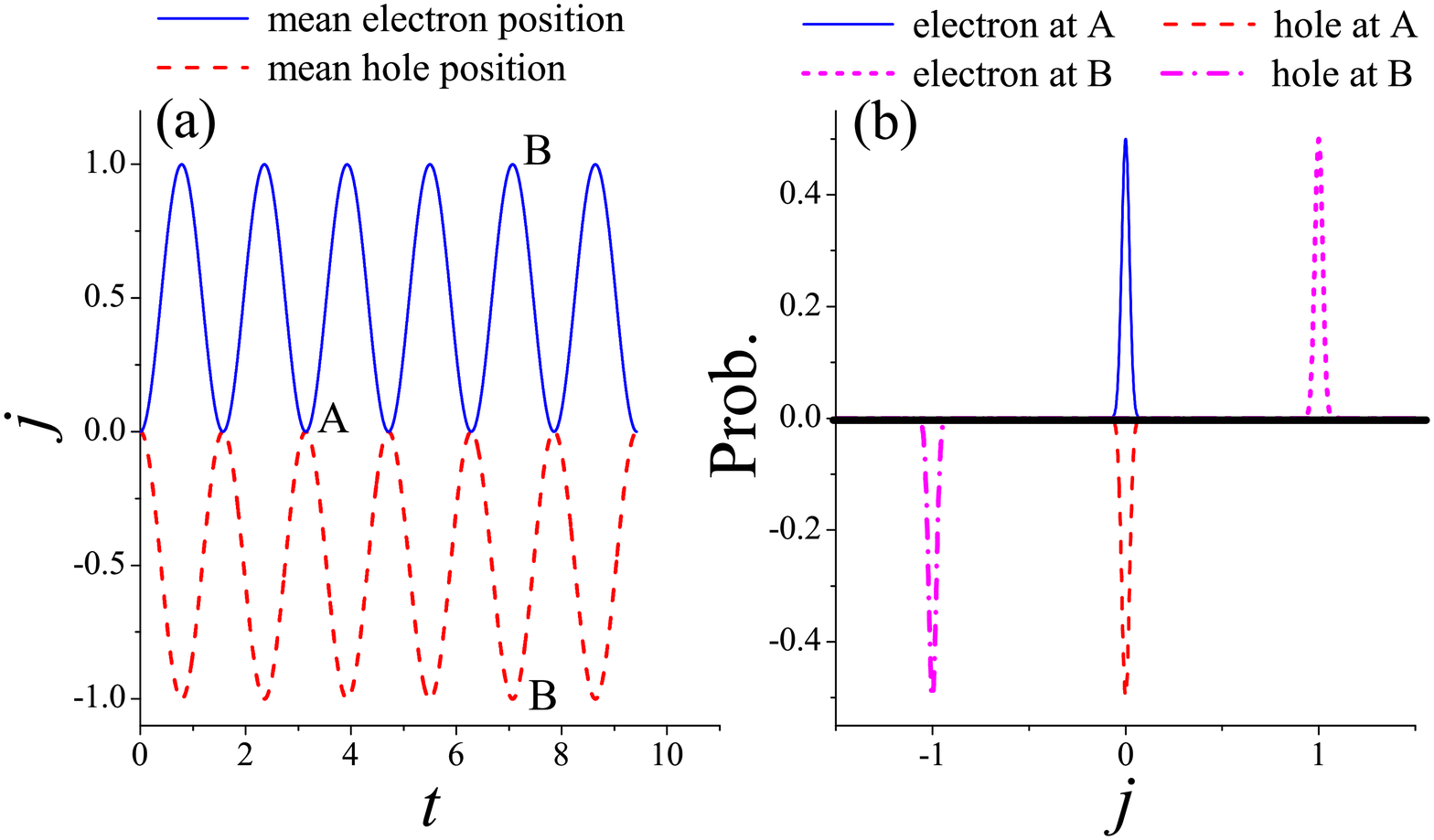}}
\end{center}
\par
\vspace*{-0.5cm} \caption{(Color online) Analytical Fourier transformations plus numerical simulations of wavepacket dynamics showing ZB oscillations in the Kitaev chain model, with the system parematers chosen to be $\mu=0$, $t_p=d$. The initial wavepackets of the particle and of the hole on top of each other localize on one lattice site, with its width determined by the lattice potential.
(a) the ZB oscillations for particle and hole as shown in the time dependence of their mean position; (b) the wavepacket profile for both particle and hole components, at time points indicated by A and B in (a). The negative occupation stands for the hole state.  The position parameter $j$ is in units of the lattice constant and $t$ is in units of $\hbar/d$. }
\end{figure}

\section{Perfect ZB oscillations subject to on-resonance driving}

We have shown that the ZB oscillations can be made perfect by choosing the right system parameter in the Kitaev Hamiltonian.  During each period of oscillation, the particle wavepacket and the hole wavepacket can be separated even though they are on top of each other at time zero.  However, as suggested by our theory above and by our numerical experiments, the amplitude of such ideal ZB oscillations is one lattice site only. It would be interesting to further convert such ZB oscillations to a more dramatic effect, which may be of experimental relevance in understanding and probing the system from a novel perspective.

Our early work suggested that on-resonance modulation of a ZB Hamiltonian can convert ZB oscillations into directed motion \cite{EPLzhang}. On a lattice, both the phase and the magnitude of the tunneling paramater $t_p$ may be modulated by introducing a high-frequency driving field \cite{njp}.
Consider then what happens if the sign of $t_p$ (we also assume $\mu=0$ in this section) is reversed after every half period of ZB, i.e., after every $T/2$. The $t_p$ parameter is hence modulated at precisely the same ZB frequency.
The Hamiltonian for the second time interval of duration $T/2$ is hence given by
\begin{equation} \label{effective1}
H_{\text{eff}}=2d \left(\begin{array}{cc}-\cos(k)& {\rm i}\sin(k)\\ -{\rm i}\sin(k)&\cos(k)\end{array} \right),
\end{equation}
with its initial state  being $|\psi_k(T/2)\rangle$. This state evolves from the first interval of duration $T/2$ and its specific form is already given in Eq.~(\ref{ipkf}), now with $\mu=0$ and $t_p=d$ using our parameter choice. The switch of the sign of $t_p$ leads to a reversal of the ZB oscillation because effectively, the field directions experienced by the particle and by the hole are exchanged.  As such, after this sign switch, the particle wavepacket will now evolve in precisely the same manner as how the hole wavepacket would evolve in the absence of the sign switch, and  the hole wavepacket will now evolve in precisely the same manner as how the particle wavepacket would evolve in the absence of the sign switch.
Therefore, at the end of the second time interval of duration $T/2$, the state becomes
\begin{equation} \label{ipkff}
|\psi_k(T)\rangle=\frac{1}{\sqrt{2}}\left(\begin{array}{c}g(k)\\g(-k)\end{array}\right)e^{-2{\rm i}k},
\end{equation}
By use of the inverse Fourier transformation, the first component of the state in Eq.~(\ref{ipkff}) is seen to represent a particle wavepacket centered at $j=2$, whereas the second component of the state in Eq.~(\ref{ipkff}) is seen to stand for a hole wavepacket centered at $j=-2$. Interestingly, the net result after duration $T$ is a particle-hole separation of four lattice sites.

Repeating this strategy, i.e., changing the signs of $t_p$
after every interval of $T/2$, the direction of the perfect ZB oscillation is consecutively reversed after each half oscillation period.  Then the particle and hole wavepackets are separated more and more, with each period $T$ contributing an increase of $4$ sites in the separation.  Of course, because this modulation scheme replies on the oscillation phases, the time when the modulation starts can also make a difference.

We have carried out numerical experiments to confirm these insights. Figure 3 presents details of the wavepacket dynamics if the sign of $t_p$ is modulated for certain periods, switched off, and then switched on again.  In case (a), the on-resonance modulation is switched off for several periods of ZB and then it is on again.  First, the particle-hole separation grows linearly with time, then it oscillates around a constant value because the modulation is off, and finally
 it grows linearly again.  In case (b), the modulation is off for a multiple ZB period plus one half ZB period.  In this case, once the modulation is switched on again, the particle and hole separation starts to decrease linearly with time and can return to zero.  In both cases, the wavepacket profile remain Gaussian all the time.

\begin{figure}[t]
\begin{center}
\vspace*{-0.5cm}
\par
\resizebox *{8.8cm}{9cm}{\includegraphics*{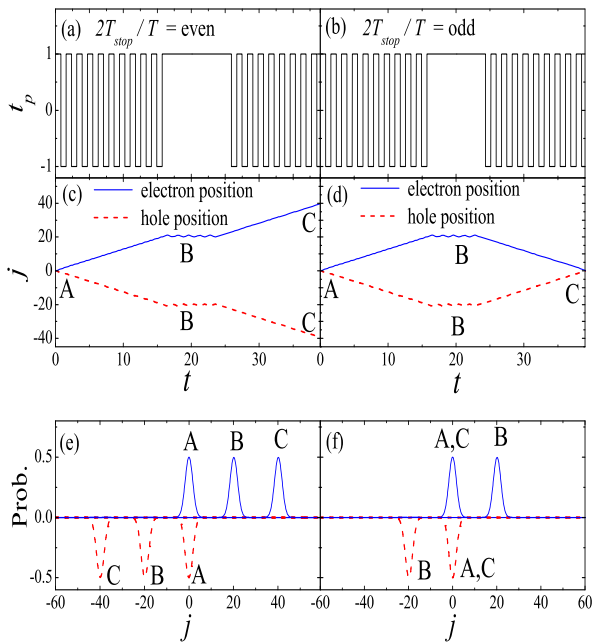}}
\end{center}
\par
\vspace*{-0.5cm} \caption{(color online) Numerical results of wavepacket dynamics in the Kitaev chain if the system parameter $t_p$ is periodically modulated.
 Panels (a) and (b) illustrate two modulation
schemes.  When
modulation is on, the sign of $t_p$ is reversed after each time
interval of $T/2$ ($T$ is the ZB period). The
modulation off-period $T_\text{stop}$ is an integer multiple of
$T$ in (a) and a half-integer multiple of $T$ in
(b).   Panels (c) and (d) depict the time-dependence of the mean position ($j$) of the particle and the hole wavepackets,
 for modulation schemes in
(a) and (b), respectively. It is seen that in (c), the particle-hole separation
continues to grow linearly after the modulation is switched on
again; but in
 (d), the change in the particle-hole separation is reversed.
 Panels (e) and (f) show the wavepacket profile, where the negative occupation stands for the hole component, at three different time points denoted by point A, B and C, for modulation schemes in (a) and (b), respectively.
 Note that in (f), the wavepacket at C is on top of that at A. $t_p$ is in units of $d$, $t$ in units of $\hbar/d$ and $j$ in units of lattice constant $a$.} \label{fig2}
\end{figure}

In all previous models simulating the ZB physics, ZB oscillations are often qualitatively understood in terms of a quantum coherence effect between two spin components, which requires the interference between two spin components and hence requires their spatial wavefunctions to overlap with each other. However, one interesting observation made from Fig.~3 is as follows.  After the particle and hole wavepackets have separated completely [see point B and the small oscillations during which the modulation is off, in both Fig.~3(c) and Fig.~3(d)], the perfect ZB oscillations (without any parameter modulation) still persist.   This is markedly different from previously studied ZB oscillations.

Let us now explain this observation.  According to the insights offered by two previous ZB studies \cite{cold2,zhang}, the ZB ocillation amplitude is the largest if the initial spinor is perpendicular to the effective ``magnetic field" depicting $H_{\text{eff}}$ [see Eq.~(\ref{effective}) or Eq.~(\ref{effective2})]. The ``magnetic field" in our model is in the ``$y-z$" plane, and indeed we have chosen the initial spinor state parallel to the ``$x$" direction, i.e., $a=\pm b=\frac{1}{\sqrt{2}}$, to get the largest ZB oscillation amplitude (which is exactly one lattice site). Consider next one particle wavepacket and one hole wavepacket separated in real space by $2D$ lattice sites, i.e.,
\begin{equation}
\langle j|\psi_{\text{electron}}\rangle=\frac{1}{\sqrt{2}}G(j-D),
\end{equation}
\begin{equation}
\langle j|\psi_{\text{hole}}\rangle=\frac{1}{\sqrt{2}}G(j+D).
\end{equation}
In the momentum space,
their respective wavefunctions will be given by $\frac{1}{\sqrt{2}}g(k)e^{-iD}$ and $\frac{1}{\sqrt{2}}g(k)e^{iD}$. Now, if the effective ZB Hamiltonian couples the two wavefunction components at the same momentum (as in previous ZB Hamiltonians), then the corresponding spinor would become
\begin{equation}
\left(\begin{array}{c}
\frac{1}{\sqrt{2}}g(k)e^{-iD}  \\ \frac{1}{\sqrt{2}}g(k)e^{iD}
\end{array}\right),
\end{equation}
which  no longer represents a pseudo-spinor wavefunction lying in the ``$x$" direction.
Then the ZB oscillations afterwards would be suppressed if $D$ is not small. By contrast,
in our model $H_{\text{eff}}$ couples $k$ and $-k$ components, so the actual spinor in the representation of $H_{\text{eff}}$ is
\begin{equation}
\left(\begin{array}{c}
\frac{1}{\sqrt{2}}g(k)e^{-iD}  \\ \frac{1}{\sqrt{2}}g(-k)e^{-iD}
\end{array}\right) =
\frac{1}{\sqrt{2}}g(k)e^{-iD} \left(\begin{array}{c} 1 \\ 1
\end{array}\right),
\end{equation}
which stays in the ``$x$" direction for arbitrary $k$. This enhances our understanding
of why ZB oscillations here sustain a complete separation between the particle and hole wavepackets.

\section{Summary}

In a typical model describing low-temperature superconductivity, there always exists a coupling to induce the pairing of electron and hole of opposite momentum values.  As we have shown based on the Kitaev model, such kind of coupling offers an interesting mechanism for ZB oscillations.  By choosing appropriate system parameters, we have shown that the ZB oscillations can be perfect: they can last long without any amplitude damping and can perfectly maintain the spatial profile of an initial wavepacket.  The possibility of perfect ZB oscillations may provide an alternative opportunity to study superconductor models.  Furthermore, we have also shown that, by periodically modulating the tunneling parameter in resonance with the ZB oscillations, the ZB oscillations can be converted to net drifting of particle and hole along opposite directions.
We also expect our finding to be relevant to other contexts where the language of Bogoliubov quasi-particle excitation still applies.

Q. Z. thanks Erhai Zhao for discussions on the Kitaev chain and the partial support by AFOSR FA9550-12-1-0079 as a visiting scholar at George Mason University.

\end{document}